\documentclass[preprint,prb,aps,floatfix]{revtex4}
\usepackage{graphicx}
\usepackage{amssymb}
\usepackage{amsmath}
\usepackage{epstopdf}
\usepackage{bm}
\newcommand{\bb}{\begin{equation}}
 \newcommand{\en}{\end{equation}} 
 
\begin{document}
 
 \title{Effective Viscosity of a Dilute Suspension of Membrane-bound Inclusions}
 \author{Mark L. Henle$^1$ and Alex J. Levine$^{1,2}$}
 \affiliation{$^1$Department of Chemistry and Biochemistry, University of California, Los Angeles, CA 90095\\
$^2$ California Nanosystems Institute, University of California, Los Angeles, CA 90095\\}

 \date{\today}

 \begin{abstract}

 When particulate suspensions are sheared, perturbations in the shear flows around the rigid particles increase the local energy dissipation, so that the viscosity of the suspension is effectively higher than that of the solvent.    For bulk (three-dimensional) fluids, understanding this viscosity enhancement is a classic problem in hydrodynamics that originated over a century ago with Einstein's study of a dilute suspension of spherical particles.~\cite{Einstein1}   In this paper, we investigate the analogous problem of the effective viscosity of a suspension of disks embedded in a two-dimensional membrane or interface.  Unlike the hydrodynamics of bulk fluids, low-Reynolds number membrane hydrodynamics is characterized by an inherent length scale generated by the coupling of the membrane to the bulk fluids that surround it.  As a result, we find that the size of the particles in the suspension relative to this hydrodynamic length scale has a dramatic effect on the effective viscosity of the suspension.  Our study also helps to elucidate the mathematical tools needed to solve the mixed boundary value problems that generically arise when considering the motion of rigid inclusions in fluid membranes.

 \end{abstract}
 
  \maketitle

 \section{Introduction}
 
  The dynamics of particulate suspensions in a viscous fluid are central to a variety of fundamental scientific questions in hydrodynamics, soft condensed matter, and biological physics.  A rather common and useful simplification of these studies replaces this heterogeneous system with a coarse-grained homogeneous one that has modified physical parameters such as viscosity.  In bulk (three-dimensional) suspensions, understanding this change in viscosity at a quantitative level has captured the interest of researchers in various disciplines for over a hundred years. Beginning with Einstein,~\cite{Einstein1, Einstein2, Einstein3} the basic physical interpretation of this result emerged:  Under an externally imposed shear, the fluid in the absence of the suspension adopts a spatially uniform shear stress and dissipates energy per unit volume proportional to that stress.  The coefficient of proportionality is the bulk fluid viscosity $\eta_{\rm 3D}$.  With the addition of the particulate suspension this uniform shear stress becomes incompatible with the flow boundary conditions at the surfaces of the particles, leading to more complex flows surrounding the particles.  These additional flows cause additional energy dissipation in the fluid.  Thus, the coarse-grained homogeneous fluid must have a \emph{higher} viscosity than the original fluid. This \emph{effective} viscosity $\eta^{\rm eff}_{\rm 3D}$ must depend on the particulate volume fraction: The larger the volume fraction, the more energy dissipated by the suspension and thus the higher the effective viscosity.  For a dilute suspension of spherical particles of radius $a$ and number density $n$, Einstein~\cite{Einstein1} found that, to leading order, the effective viscosity depends on the volume fraction $\phi_{\rm 3D} = \frac{4}{3} \pi a^3 n$ as
 \bb
 \label{eq:Eta3Deff}
 \eta_{\rm 3D}^{\rm eff} = \eta_{\rm 3D} \left[1+ \frac{5}{2} \phi_{\rm 3D}\right].
 \en
 This result has been extended to non-rigid droplets in a fluid~\cite{Taylor} and even non-spherical geometries,~\cite{Happel, Hinch} where changes in the numerical prefactor are found.   In all these cases, the results apply only at low particulate volume fractions. Experiments find that Eq.~(\ref{eq:Eta3Deff}) holds for $\phi_{\rm 3D} \lesssim 0.01$.~\cite{Kynch, Manley}  Above these concentrations the hydrodynamic interactions between particles, which are neglected in these calculations, become important.  At such volume fractions one must consider these effects, as well as the possibility that the imposed shear flow changes the microstructure of the suspensions.~\cite{Happel, Brady1, Segre}

 In contrast to this tremendous effort in exploring the effect of finite particulate concentrations on the viscosity of \emph{three-dimensional} suspensions, comparatively little is known about the analogous problem for fluid membranes and interfaces.  The problem of membrane hydrodynamics is complicated by the interactions of the essentially two-dimensional viscous membrane with the surrounding three-dimensional solvents.  Because of this coupling, in-plane fluid momentum in the membrane is not conserved:  Around a moving point-like particle in the membrane, momentum transfers to the surrounding fluids over a length scale set by the ratio of the 2D membrane viscosity $\eta_{\rm m}$ to the 3D solvent viscosity $\eta_{\rm 3D}$.  This \emph{Saffman-Delbr\"{u}ck length} $\ell_0 \sim \eta_{\rm m}/ \eta_{\rm 3D}$~\cite{Saffman1, Saffman2} makes membrane hydrodynamics qualitatively distinct from the usual three-dimensional hydrodynamics of bulk liquids, since the latter theory has no analogous inherent length scale (in the limit of vanishing Reynolds number).  
 
 The existence of an inherent length scale in membrane hydrodynamics has profound implications on the transport properties of membranes.  The mobility $\mu$ of a particle in an overdamped bulk 3D fluid is always inversely proportional to its size $a$, $\mu \sim 1/a$,~\cite{Landau}  as long as the Reynolds number remains small.  In contrast, the mobility of a membrane-bound inclusion exhibits two drastically different behaviors as the particle size is varied, depending on the ratio $a / \ell_0$.  When the particle size is small compared to $\ell_0$, the flows in the membrane dissipate much more energy than those in the surrounding bulk fluids, and the mobility only has a weak logarithmic dependence on the particle size.~\cite{Saffman1, Saffman2, Hughes, Levine1}  Conversely, when the particle size is large compared to $\ell_0$, the flows in the bulk dissipate more energy.  Not surprisingly, this leads to a mobility that, like its three-dimensional counterpart, is inversely proportional to $a$,~\cite{Hughes} although the numerical prefactor is different.   This complex dependence of mobility on particle size, as well as the related complex distance-dependence of hydrodynamic interactions,~\cite{Levine1, Levine2, Levine3, Stone1} have been strongly supported by several experiments.~\cite{Klingler, Cicuta, Prasad}  However, we point out that recent work has suggested that protein transport in lipid bilayers is more subtle than suggested by the original Saffman-Delbr\"{u}ck model.~\cite{Gambin, Guigas, Naji}  In spite of this subtlety, understanding the effective membrane viscosity remains important for studies of the diffusive properties of transmembrane proteins.  
  
 In this article we examine the effect of a finite but small concentration of membrane-bound particles on the effective membrane viscosity.  In essence we wish to find a relation analogous to Eq.~(\ref{eq:Eta3Deff}) expressing the dependence of the effective membrane viscosity on the \emph{area fraction} of membrane inclusions.   We have several different motivations to study this problem.  First, there is the fundamental question of how the Saffman-Delbr\"{u}ck length enters the coefficient of the area fraction term in the membrane version of Eq.~(\ref{eq:Eta3Deff}).  As with the mobility of a membrane-bound inclusion, that coefficient should be a function of the dimensionless ratio $\epsilon \equiv a/\ell_0$ for particles of radius $a$. Secondly, the effect of particulate suspensions on membrane viscosity addresses important biophysical questions regarding the dynamics of proteins embedded in the plasma membrane of cells.  It is now well known that cell membranes are crowded environments in which the diffusive transport of transmembrane proteins controls, for example, cell adhesion and cell-cell signaling.  These problems in protein dynamics are currently the focus of much study.~\cite{Schwartz}  While much attention has been paid to how immobile obstacles in the membrane can hinder diffusion,~\cite{Kusumi, Saxton} comparatively little has been paid to how a suspension of mobile particles can have a similar effect  by increasing the viscosity of the membrane.  
 
 Finally, this study provides a simple physical system in which to explore a class of complex mathematical problems known as \emph{dual integral equations}.  Such problems arise generically in systems involving the transport of rigid inclusions in fluid membranes.  The fundamental mathematical difficulty presented in these systems is that their behavior is governed by the solution to a \emph{mixed boudary-value problem}.  Physically, this arises from the two distinct regions in the system:  The fluid regions of the membrane and the solid regions of the particle interior.  In the former region, a stress continuity condition applies; that is, the internal stresses caused by the flows in the membrane must be balanced by the external stresses exerted on the membrane by the surrounding solvents. In the interior of the inclusion, on the other hand, the particle's rigidity supplies arbitrary constraint stresses to ensure that the entire inclusion executes only rigid body motions.   As a result, the boundary condition in this region becomes a ``stick'' velocity balance condition.  Each of these boundary conditions is expressed as an integral equation, so that the system requires two separate integral equations to be satisfied simultaneously in two non-overlapping domains of the membrane. Dual integral equations such as these have certainly been studied before,~\cite{Hughes, Stone2, Sneddon, Tranter1, Tranter2}  even in the context of membrane hydrodynamics,~\cite{Hughes, Stone2} but we believe that the methods are not widely known.  To that end, we include an explanatory Appendix recapitulating the basic mathematical tools needed to solve these dual integral equations as they arise in membrane hydrodynamics.  By mastering these tools here, we open the possibility of exploring numerous related problems, including the lubrication forces between two approaching transmembrane proteins or lipid rafts. 
 
To briefly summarize our results, we find that, like the mobility of a membrane-bound inclusion, the effective membrane viscosity divides naturally into two regimes, characterized by the value of the dimensionless parameter $\epsilon = a/\ell_0$.  When $\epsilon \ll 1$, we find that the effective viscosity does indeed behave in a manner analogous to the three-dimensional result Eq.~(\ref{eq:Eta3Deff}); that is, we find $\eta_{\rm m}^{\rm eff} = \eta_{\rm m} [1+ f(a/\ell_0) \phi]$, where the coefficient of the area density term is a function of $\epsilon$.  When $\epsilon \gg 1$, however, most of the dissipation occurs in the bulk solvents, so it is more appropriate to consider the inclusions as shifting the viscosity of the three-dimensional viscosity of these solvents.  Here, we find a result nearly identical to the original Einstein result Eq.~(\ref{eq:Eta3Deff}). In addition, our calculation provides a solution for the effective viscosity for arbitrary $\epsilon$ that interpolates between these regimes.  The full solution relies on numerical integration and matrix inversion, but we provide a closed form approximate solution that is exact in both asymptotic limits and has at most a small ($< 10\%$) error for intermediate values of $a/ \ell_0$.  

The remainder of the paper is organized as follows:  In Section~\ref{sec:flows}, we determine the shear flows and dynamic pressures around an isolated membrane inclusion.  These flows are then used in Section~\ref{sec:EffectiveViscosity} to determine the effective viscosity of a dilute suspension of such inclusions.  We conclude in Section~\ref{sec:Conclusion} with a summary of our results and a discussion of problems in the hydrodynamics of rigid membrane-bound inclusions which we plan to address in the future using the knowledge we have gained here.

\section{Isolated Inclusion}
\label{sec:flows}

\begin{figure}
 \includegraphics[scale=1.1]{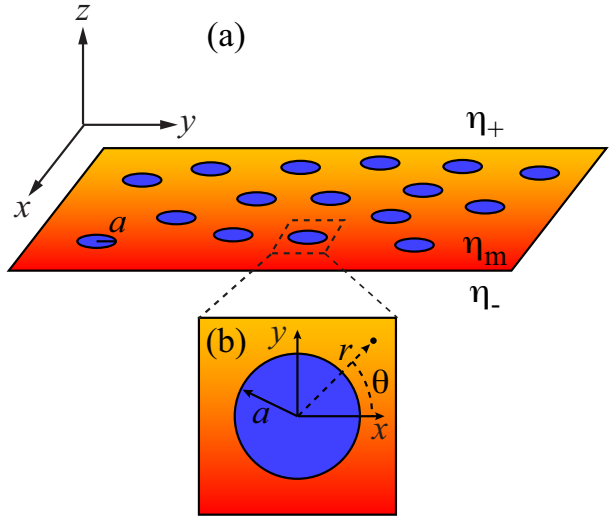}
 \caption{\label{fig:Schematic1} (a) Schematic illustration of a membrane (viscosity $\eta_{\rm m}$) containing a suspension of disks (radius $a$) and surrounded by bulk fluids above ($z>0$, viscosity $\eta_+$) and below ($z<0$, viscosity $\eta_-$). (b) Detail of an isolated inclusion in the membrane, viewed from above.  The origin of the in-plane Cartesian $(x,y)$ and cylindrical $(r, \theta)$ coordinates is the center of the inclusion.  }
 \end{figure}

Consider a flat, two-dimensional membrane (located at $z=0$) consisting of a distinct fluid of viscosity $\eta_{\rm m}$.  The membrane is surrounded by bulk  fluids above ($z>0$) and below ($z<0$) whose shear viscosities are $\eta_+$ and $\eta_-$, respectively; see Figure~\ref{fig:Schematic1}(a).  We assume that all three fluids are incompressible and that all flows occur at low Reynolds number.  Thus, the membrane velocity field $\mathbf{v}^{\rm m}$ must obey the 2D incompressible Stokes equation:
\bb
\label{eq:2DStokes}
\eta_{\rm m} \nabla_\perp^2 v_\alpha^{\rm m} + \left. \sigma_{\alpha z}^+ \right|_{z=0}- \left. \sigma_{ \alpha z}^- \right|_{z=0}- \partial_\alpha p^{\rm m}=0,
\en
\bb
\label{eq:2DIncompress}
\mathbf{\nabla} \cdot \mathbf{v}^{\rm m}=0,
\en
while the velocity fields of the bulk fluids $\mathbf{v}^\pm$ obey the incompressible 3D Stokes equation:
\bb
\label{eq:3DStokes}
\nabla^2 \mathbf{v}^\pm = -\mathbf{\nabla}P^{\pm}, \qquad \mathbf{\nabla} \cdot \mathbf{v}^\pm = 0.
\en
Here, $p_{\rm m}$ and $P_\pm$ are the membrane and bulk fluid pressures respectively, and $\sigma_{ij}^\pm = \eta_\pm \left[ \partial_i v_j^\pm +  \partial_j v_i^\pm\right]$ is the bulk fluid stress tensor.  

In general, any membrane flow field can be decomposed into three linearly independent normal modes, which correspond to the compression, bending, and shearing of the membrane.  The out-of-plane bending deformations are decoupled at linear order from the in-plane flows.  Since the focus of this paper is the dissipation caused by the in-plane flows, we ignore all bending deformations. Furthermore, we eliminate the compression modes by our assumption of the incompressibility of the membrane.  This assumption is generally appropriate for lipid bilayers. Thus, the hydrodynamic flows in the membrane can be decomposed purely into shear modes; that is, any membrane fluid velocity field can be written as a linear superposition of these modes.  It is known that pure shear flows in a flat membrane generate no pressure in the surrounding bulk fluids,~\cite{Levine1} so we set $P^\pm =0$.

In order to calculate the effective viscosity of the membrane, we follow loosely the derivation of Einstein's result for the effective viscosity of a dilute three-dimensional colloidal suspension given in Ref.~\onlinecite{Landau}.  We impose a simple shear flow in the absence of the particulate suspension and calculate the resultant dissipative stress in the system.  These flows in the membrane and surrounding solvents act as a probe of the viscous dissipative processes in the system.  We then add a single rigid particle to the membrane and calculate the consequent perturbation to the flow fields.  Using these flows, we calculate the average stress tensor in a dilute suspension of such particles in the membrane.  By examining the terms that arise from the in-plane dissipative flows, we extract the effective viscosity.

The simplest membrane shear flows generate constant (i.e. position-independent) stresses.  Thus, we choose the ``unperturbed'' membrane velocity $\mathbf{v}_0^{\rm m}$ to be of the form
 \bb
 \label{eq:v0M}
 v_{0, i}^{\rm m} (x, y) = \delta_{\alpha i}^\perp c_{\alpha \beta} x_\beta,
  \en
 where $c_{\alpha \beta}$ is a traceless symmetric tensor.  Throughout this paper we use Greek indices for the in-plane (2D) vector component $x,y$ and Latin indices for the 3D vector components $x,y,z$; the delta function $\delta_{\alpha i}^\perp$ projects the Latin indices onto the Greek indices.   The symmetry of the tensor [$c_{\alpha \beta}=c_{\beta \alpha}$] excludes flows corresponding the rigid rotation of the entire membrane. Such flows generate no dissipative stresses in the membrane and therefore are unnecessary.  The vanishing trace [$c_{\alpha \alpha}=0$] enforces the incompressibility constraint Eq.~(\ref{eq:2DIncompress}).  
 
Given the velocity field Eq.~(\ref{eq:v0M}), we need to determine the bulk fluid flows $\mathbf{v}_0^\pm$ and membrane pressure $p_0^{\rm m}$. The bulk flows are governed by the incompressible Stokes equation, Eq.~(\ref{eq:3DStokes}).  The boundary conditions are given by the usual ``stick'' boundary conditions  at the surface of the membrane, $ \mathbf{v}_0^\pm (x, y, 0) = \mathbf{v}_0^{\rm m} (x, y)$, as well as the the 2D Stokes equation, Eq.~(\ref{eq:2DStokes}).  It is straightforward to show that the shear flow in the membrane Eq.~(\ref{eq:v0M}) induces the same shear flows in the bulk fluids:
 \bb
 \label{eq:v0}
 v_{0,i} (x,y, z) = \delta_{i \alpha}^\perp c_{\alpha \beta} x_\beta.
 \en
Here and throughout the paper, we use the vector field $\mathbf{v}(x,y,z)$ to represent the velocity field throughout all space:
 \bb
 \mathbf{v} (x,y,z) \equiv
 \begin{cases}
 v^- (x,y,z) & z<0\\
 v^{\rm m} (x,y) & z=0\\
 v^+ (x,y,z) & z>0\\
 \end{cases}.
 \en
 Finally, the membrane pressure vanishes for the unperturbed flows, $p_0^{\rm m}=0$.  This solution satisfies Eqs.~(\ref{eq:2DStokes})-(\ref{eq:3DStokes}), as well as the stick boundary conditions, and is thus the unique solution for the velocity field at all points in the system.  

We now introduce an isolated membrane inclusion, a rigid disk of radius $a$ and of negligible thickness, at the origin of our coordinate system; see Fig.~{\ref{fig:Schematic1}(b).  Its presence perturbs the flows in the system and introduces new boundary conditions not satisfied by the unperturbed flows given above.  Due to the linearity of the Stokes equation, we can write the total fluid velocity as $\mathbf{v} = \mathbf{v}_0 + \mathbf{v}_1$; that is, $\mathbf{v}_1$ is the ``perturbative'' flow field caused by the inclusion.  It is clear from the rotational symmetry of Eq.~(\ref{eq:v0}) that the disk remains at rest:
\bb
\label{eq:v1BC1}
\mathbf{v}_1^{\rm m} (r, \theta) = -  \mathbf{v}_0^{\rm m} (r, \theta)  = -  r c_{\alpha \beta} n_\beta \qquad r \leq a,
\en
where $r, \theta$ are the radial and angular variables, respectively, in cylindrical coordinates [see Fig.~\ref{fig:Schematic1}(b)], and $n_\alpha \equiv x_\alpha/r$ is the in-plane unit vector.  Furthermore, the perturbative flows must vanish far away from the disk:
\bb
\label{eq:v1BC2}
\lim_{r \rightarrow \infty} \mathbf{v}_1 (r, \theta,z)= \lim_{z\rightarrow \pm \infty}\mathbf{v}_1 (r, \theta, z) = 0.
\en
  The final boundary condition is given by the 2D Stokes equation, Eq.~(\ref{eq:2DStokes}), which holds everywhere outside of the disk $r>a$.  We note that if we had included rotational flows in the unperturbed membrane flows (i.e. if we allowed $c_{\alpha \beta}$ to have antisymmetric parts) then the disk would simply rotate rigidly with the fluid, thus generating no additional sources of dissipation.

Since we have chosen the unperturbed membrane flows $\mathbf{v}_0^{\rm m}$ to be entirely composed of shear modes, and these modes are linearly independent of the bending and compression modes of the membrane, the perturbative velocity field $\mathbf{v}_1^{\rm m}$ must also consist solely of shear modes.  As a result, it can be written as an antisymmetric derivative of a scalar field:
\bb
\label{eq:v1M}
 v_{1,\alpha}^{\rm m} (r, \theta) = \epsilon_{\alpha \beta} \partial_\beta \psi_1 (r,\theta),
 \en
where $\epsilon_{\alpha \beta}$ is the antisymmetric unit tensor.  Indeed,  $\mathbf{v}_0^{\rm m}$ can also be written in this form, with the scalar field
\bb
	\psi_0 (r, \theta)= -\frac{r^2}{2}\left[c_{xy} \cos 2\theta- c_{xx}\sin 2\theta \right] \equiv -r^2 \Theta (\theta).
\en

We know from the linearity of the Stokes equation and the azimuthal symmetry of the disk that the angular dependence of $\mathbf{v}_1^{\rm m}$ is set by the boundary condition Eq.~(\ref{eq:v1BC1}). Thus, the angular dependence of $\psi_1 (r, \theta)$ must be identical to that of $\psi_0 (r, \theta)$: $\psi_1 (r, \theta) =  \rho (r) \Theta (\theta)$.  Using separation of variables, it is straightforward to show that the incompressibility constraint in Eq.~(\ref{eq:3DStokes}) and the stick boundary condition at the membrane surface $z=0$ imply that the bulk fluid velocities $v_{1,i}^\pm (r, \theta, z)$ have the form
\bb
v_{1,i}^\pm (r, \theta, z) = \delta_{\alpha i}^\perp h^\pm (z) \epsilon_{\alpha \beta} \partial_\beta \psi_1 (r,\theta).
\en
 Then the Stokes equation for the bulk velocities, Eq.~(\ref{eq:3DStokes}), becomes
\bb
\label{eq:v1Stokes}
\frac{\rho''(r)+\frac{1}{r}\rho'(r)-\frac{4}{r^2}\rho(r)}{\rho (r)} = -\frac{h^{\pm''} (z)}{h^\pm(z)}=const.
\en
From the boundary condition Eq.~(\ref{eq:v1BC2}), it is clear that we should choose exponential decays for $h(z)$.  Then the solution to Eq.~(\ref{eq:v1Stokes}) is given by
\bb
\label{eq:PsiOneBulk}
\psi_1 (r, \theta) h^\pm (z) =  a^2 \Theta (\theta) \int_0^\infty \frac{dq}{q} \, B(q) J_2 (q u) e^{- k \zeta},
\en
where $u \equiv r/a$, $\zeta \equiv \left| z \right|/a$, and  $J_2 (q u)$ is the second order Bessel function of the first kind.  The Bessel function of the second kind $Y_2 (q u)$ is also a solution to Eq.~(\ref{eq:v1Stokes}),but it fails to satisfy the requirement of finite fluid velocities at $r=0$.  

The function $B(q)$ is a \emph{modified Hankel transform} of the function $\rho (u)$, i.e. the radial dependence of the scalar field $\psi_1$.  In general, the kernel of these transforms is the product of a Bessel function $J_m (q u)$ with $q^p$ for arbitrary real numbers $m$ and $p$.  Using the orthogonality and completeness of the Bessel functions, it can be shown that there is a one-to-one mapping of the function $\rho (u)$ defined on the half-line $0<u<\infty$ and its modified Hankel transform $B(q)$.~\cite{Arfken}   At the moment, $B(q)$ is an undetermined function.  Its form is determined by the boundary conditions in the membrane, which are given below. 

It is straightforward to show using Eq.~(\ref{eq:PsiOneBulk}) that 
\begin{align}
\label{eq:v1}
\notag
v_{1,i} (r, \theta, z) = \frac{a}{2} &\delta_{\alpha i}^\perp \int_0^\infty dq \, B(q) e^{-q \zeta} \Bigg[- J_3 (q u) c_{\mu \nu} n_\alpha n_\mu n_\nu\Bigg.\\
&\Bigg. +\left(J_3 (q u)-\frac{2 J_2 (q u)}{q u}\right) c_{\alpha \beta} n_\beta \Bigg].
\end{align}

In order to determine the function $B(q)$, we need to enforce the boundary conditions in the membrane.  These boundary conditions are integral equations for $B(q)$, because the velocity Eq.~(\ref{eq:v1}) is itself an integral equation.  Inside the disk -- that is, for $u<1$ -- we impose the stick boundary condition Eq.~(\ref{eq:v1BC1}).   Since this condition must be satisfied for arbitrary $\theta$, we can see from Eq.~(\ref{eq:v1}) that we obtain two separate integral equations:
\bb
\label{eq:IntEqIn}
\int_0^\infty dq \, q^{-1} B(q) J_2 (q u) = u^2, \qquad u<1,
\en
\bb
\label{eq:IntEqIn2}
 \int_0^\infty dq \, B(q) J_3 (q u) = 0, \qquad u<1.
\en
Outside of the disk, we have the stress balance condition Eq.~(\ref{eq:2DStokes}).  By taking the anti-symmetric derivative  $\epsilon_{\alpha \mu} \partial_\mu$ of this equation, we can eliminate the membrane pressure.  Then, using Eq.~(\ref{eq:PsiOneBulk}), we obtain the final integral equation:
\bb
\label{eq:IntEqOut}
\int_0^\infty dq \, q^2 B(q) J_2 (qu) \left[1+\frac{q}{\epsilon}\right] = 0, \qquad u>1,
\en
where $\epsilon \equiv \frac{a}{\ell_0}$, with the Saffmann-Delbr\"{u}ck length $\ell_0 \equiv \frac{\eta_{\rm m}}{\eta_+ + \eta_-}$.  The parameter $\epsilon$ is the key control parameter for the hydrodynamics of membrane-bound inclusions.  When $\epsilon \gg 1$, the flows in the membrane dissipate much more energy than the induced flows in the bulk fluids; conversely, when $\epsilon \ll 1$, the dissipation occurs primarily in the bulk. 

Finally, we obtain the membrane pressure using Eqs.~(\ref{eq:2DStokes}),~(\ref{eq:v1}), and~(\ref{eq:IntEqOut}):
\begin{align}
\label{eq:p1}
\notag
p_1^{\rm m} (r, \theta) = \frac{\eta_{\rm m}}{4} c_{\alpha \beta} n_\alpha n_\beta &\int_0^\infty dq\, B(q)\left(q+\epsilon\right)\\
&\times \left[q u J_1 (q u) - 2 J_2 (q u)\right].
\end{align}

The integral Eqs.~(\ref{eq:IntEqIn})--(\ref{eq:IntEqOut}) completely determine the modified Hankel transform $B(q)$, which in turn determines the fluid velocities and pressures everywhere in the membrane and bulk fluids.  However, finding the solution to these integral equations is difficult.  The difficulty arises from the fact that this is a mixed boundary value problem:  Inside the disk ($0<u<1$), we have a Dirichlet boundary condition that sets the total membrane velocity to zero; Outside the disk, we have a Neumann boundary condition that imposes stress balance across the fluid membrane.  As a result, the boundary conditions Eqs.~(\ref{eq:IntEqIn})--(\ref{eq:IntEqOut}) ultimately reduce to a pair of \emph{dual integral equations.} Specifically, we must find the transform $B(q)$ that satisfies Eq.~(\ref{eq:IntEqIn}) inside the disk and Eq.~(\ref{eq:IntEqOut}) outside the disk;  we show in Appendix~\ref{app:DualIntegral} that Eq.~(\ref{eq:IntEqIn2}) is redundant, because it is automatically satisfied by the solution to the dual integral equations.  

By contrast, consider a problem in which the boundary condition is given by a single integral equation that is valid over the entire region $0<u<\infty$.  In this case, the integral boundary condition is easily inverted using the mutual orthogonality of the Bessel functions.~\cite{Arfken} This is analogous to the well-known inversion of the Fourier expansions of a function.  Indeed, if the size of the inclusion is very small, $a \rightarrow 0$, we can approximate it by a point-like particle and ignore the velocity balance conditions inside the disk, Eqs.~(\ref{eq:IntEqIn}) and~(\ref{eq:IntEqIn2}).  This limit, which is used often in membrane hydrodynamics,~\cite{Saffman1, Saffman2, Levine1, Levine2, Levine3} greatly simplifies the solution.  In the present problem, though, the finite size of the inclusion is essential in determining the effective viscosity of a suspension since it controls the suspension's area density. Furthermore, one of the major motivations of this study is to gain a better understanding of the mathematical difficulties encountered in solving dual integral equations.  

 The mathematical tools necessary to solve these dual integral equations are derived in Ref.~\onlinecite{Sneddon}; we summarize the necessary results in Appendix~\ref{app:DualIntegral}.  Briefly, we need to find a way to reduce the two modified Hankel transforms,  Eqs.~(\ref{eq:IntEqIn}) and~(\ref{eq:IntEqOut}), into a single modified Hankel transform valid over the entire half-line $0<u<\infty$.  Once we have accomplished this task, we can invert the remaining transform using the inverse modified Hankel transform.   In order to combine the dual integral equations, we need to transform Eqs.~(\ref{eq:IntEqIn}) and~(\ref{eq:IntEqOut}) using operators that act on these integral equations entirely within their respective regions of validity, $0<u<1$ and $1<u<\infty$.  In addition, these integral operators must possess simple convolution properties with the modified Hankel transforms.  Such operators are known as the Erd\'{e}lyi-Kober operators.~\cite{Sneddon}  
  
 In Appendix~\ref{app:DualIntegral}, we define the modified Hankel transforms and the Erd\'{e}lyi-Kober operators; we also present the relevant inversion and convolution properties of these operators.  Using these properties, the dual integral Eqs.~(\ref{eq:IntEqIn}) and~(\ref{eq:IntEqOut}) can be reduced to a single integral equation,  Eq.~(\ref{eq:SingleInt2}).  It is convenient to re-write this equation in terms of spherical Bessel functions $j_n (u) \equiv \sqrt{\pi/(2 u)} J_{n+1/2}(u)$, so that
\begin{align}
  \label{eq:SingleInt3}
   \pi \left(u+\epsilon\right) &B(u) = 16 \epsilon j_1(u)\\
  \notag &
+  \int_0^\infty dz \, z B(z)\left[j_0(u-z)-j_0(u+z)\right].
 \end{align}

Using the addition theorem for spherical Bessel functions,
\bb
j_0 (u \pm z) = \sum_{m=0}^\infty \left(2 m+1\right) j_m (u) j_m(z) \left(\mp 1\right)^{\rm m},
\en
we find that the even $m$ terms cancel, leaving
\bb
\label{eq:SingleIntSeries}
\left(u+\epsilon\right) B(u) =  \sum_{n=1}^\infty b_n (\epsilon) j_{2n-1} (u),
\en
where the coefficients are defined as
\bb
b_n (\epsilon) \equiv \frac{16 \epsilon}{\pi} \delta_{n,1} + \frac{2}{\pi}\left(4n-1\right) \int_0^\infty dz \, z B(z) j_{2n-1} (z).
\en

To solve Eq.~(\ref{eq:SingleIntSeries}), we convert it into a matrix equation for the coefficients $b_n$ by multiplying it by $u j_{2l-1} (u) /(u+\epsilon)$ and integrating.  Using the orthogonality of the spherical Bessel functions, 
\bb
\label{eq:SphBesselID}
\int_0^\infty dq \, j_{2n-1} (q) j_{2l-1} (q) = \frac{\pi \delta_{l,n}}{2(4l-1)},
\en
we find
\bb
\label{eq:bMatrixEq}
\sum_{n=1}^\infty b_n (\epsilon) \mathcal{M}_{n,l}= \frac{8}{3} \delta_{l,1},
\en
where
\bb
\label{eq:MatrixElements}
\mathcal{M}_{n,l} \equiv \int_0^\infty dq \, \frac{j_{2n-1} (q) j_{2l-1} (q)}{q+\epsilon}.
\en

Eq.~(\ref{eq:bMatrixEq}) is a matrix equation for the coefficients $b_n$; given the matrix inverse $\mathcal{M}^{-1}_{l,m}$, its solution is trivial:
\bb
\label{eq:bExact}
b_n (\epsilon) = \frac{8}{3} \mathcal{M}_{1,n}^{-1}.
\en

Given the coefficients $b_n$ the function $B(u)$ can be found from Eq.~(\ref{eq:SingleIntSeries}). Then the membrane velocity Eq.~(\ref{eq:v1}) becomes
\begin{align}
\label{eq:v1Solution}
v_{1,\alpha} (r, \theta, 0) &= \frac{a}{2} \sum_{n=1}^N b_n (\epsilon) \Bigg\{-\frac{1}{2} c_{\alpha \beta} n_\beta \mathcal{V} (u; n, 1) \Bigg. \\
\notag
&+\Bigg. \left[\frac{1}{2} c_{\alpha \beta} n_\beta -c_{\mu \nu}n_\alpha n_\mu n_\nu\right] \mathcal{V} (u; n, 3)\Bigg\},
\end{align}
where
\bb
\mathcal{V} (u;n,m) \equiv \int_0^\infty \frac{dq}{q+\epsilon} j_{2n-1}(q) J_m(q u).
\en

Finally we turn to the pressure Eq.~(\ref{eq:p1}).  It is straightforward to show using Eq.~(\ref{eq:SingleIntSeries}) that all of the integrals in the resultant expression vanish, except for the $n=1$ term in the sum [see Eq.~(\ref{eq:BesselInt})]:
\bb
\label{eq:PressureSolution}
p_1^{\rm m} (r, \theta) = -\frac{\eta_{\rm m} c_{\alpha \beta} n_\alpha n_\beta}{3u^2} b_1 (\epsilon).
\en

We can see from Eq.~(\ref{eq:v0M}) that the tensor $c_{\alpha \beta}$ has units of $[sec]^{-1}$.  Therefore, the appropriate dimensionless quantities are
\bb
\label{eq:Dimensionless}
\tilde{c}_{\alpha \beta} \equiv \tau c_{\alpha \beta}, \quad \mathbf{\tilde{v}} \equiv \frac{\tau}{a} \mathbf{v}, \quad \tilde{p}_1^{\rm m} \equiv \frac{\tau}{\eta_{\rm m}}p_1^{\rm m},
\en
where $\tau$ is the characteristic time scale in $c_{\alpha \beta}$.

\begin{figure}
%\begin{flushleft}
\includegraphics[scale=.95]{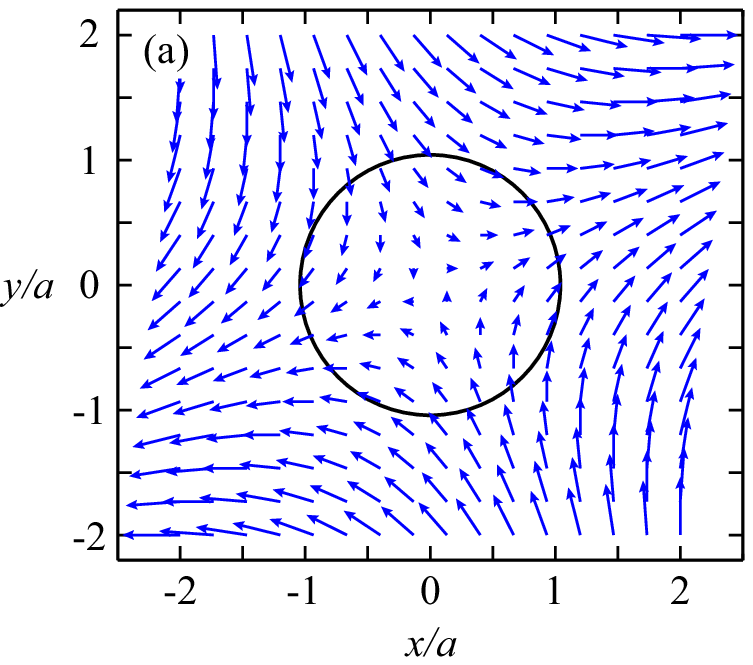}
\includegraphics[scale=.95]{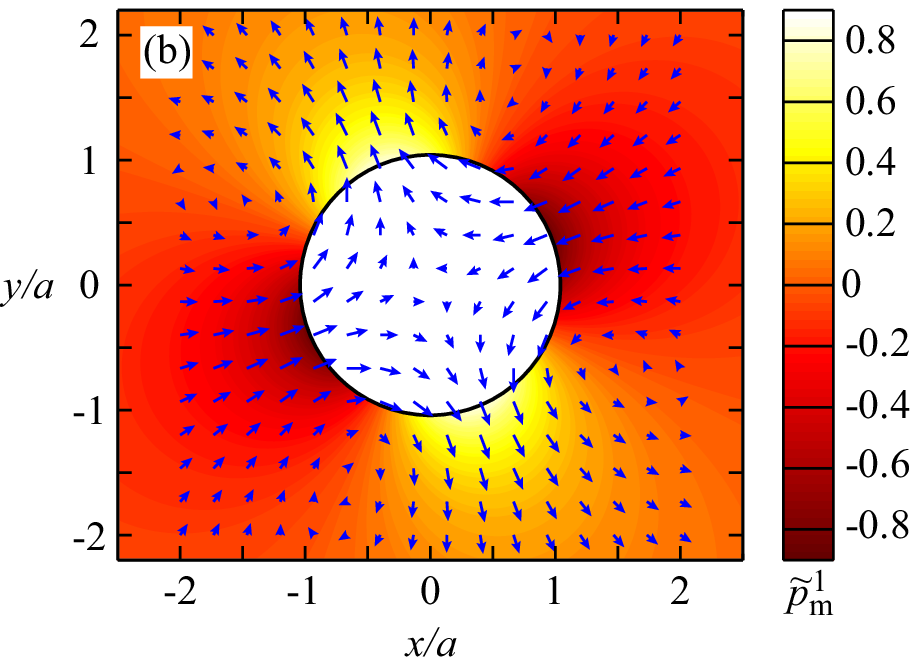}
\vspace{10pt}\\
\includegraphics[scale=.95]{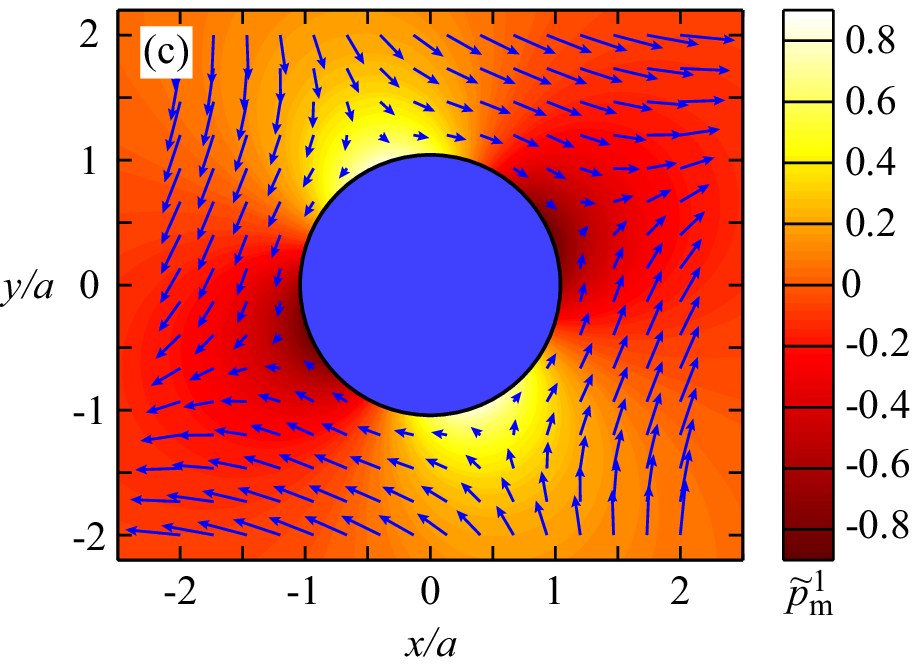}
%\end{flushleft}
\caption{\label{fig:Velocity} Dimensionless velocities [see Eq.~(\ref{eq:Dimensionless})] (a) $\mathbf{\tilde{v}_0}$, (b) $\mathbf{\tilde{v}_1}$, and (c) $\mathbf{\tilde{v}}$ around an isolated inclusion of radius $a$, as a function of the dimensionless distances $\frac{x}{a}, \, \frac{y}{a}$.  The colorfield (b, c) is the dimensionless membrane pressure $\tilde{p}_1^{\rm m}$.  All fields are calculated using $\epsilon=1$ and $\tilde{c}_{xx}=\tilde{c}_{xy}=0.1$.}
\end{figure}
Figure~\ref{fig:Velocity} shows the dimensionless membrane velocity fields $\mathbf{\tilde{v}_0}$, $\mathbf{\tilde{v}_1}$, and $\mathbf{\tilde{v}}$ and a color plot of the dimensionless pressure field $\tilde{p}_1^{\rm m}$.  In order to compute these fields, the integrals $\mathcal{M}_{n,l}$ and $\mathcal{V} (u; n, m)$ must be computed numerically, and the matrix $\mathcal{M}$ must be inverted numerically.  The details of this procedure are provided in Appendix~\ref{app:Numerics}.  Figure~\ref{fig:Velocity}(a) shows the unperturbed velocity $\mathbf{\tilde{v}_0}$, which clearly does not respect the boundary conditions at the surface of the inclusion.  The perturbative velocity field $\mathbf{\tilde{v}_1}$ shown in Fig.~\ref{fig:Velocity}(b) accounts for these boundary conditions.  We can see that the perturbative velocity inside the particle is equal and opposite to the unperturbed velocity, causing the total velocity to vanish there and thus respect the boundary condition Eq.~(\ref{eq:v1BC1}), as shown in Fig.~\ref{fig:Velocity}(c).  In addition, the insertion of the particle into the membrane gives rise to regions of positive membrane pressure where the perturbative velocity flows out of the inclusion; conversely, regions of negative membrane pressure arise where the perturbative velocity flows into the inclusion.  For Figure~\ref{fig:Velocity}, we have chosen an intermediate value of the dimensionless parameter $\epsilon = 1$, i.e. we have set $a=\ell_0$.  For different values of $\epsilon$, the velocity and pressure fields look qualitatively similar to those in Fig.~\ref{fig:Velocity}, since the boundary conditions at the surface of the inclusion must still be obeyed.  However, if we increase the viscosity of the membrane while keeping the particle size constant -- that is, if we decrease $\epsilon$ -- the gradients in the perturbative membrane velocity field $\mathbf{\tilde{v}_1}$ are decreased, causing this velocity to persist farther away from the inclusion (not shown).  In addition, the magnitude of the pressure field increases.  Conversely, higher values of $\epsilon$ lead to more localized perturbative velocity fields and smaller membrane pressures.  We can understand this behavior in the following way: As mentioned above, viscous dissipation in the membrane dominates in the limit of small $\epsilon$.  As a result, large gradients in the membrane velocity field are unsustainable, causing the perturbative velocity field at the surface of the inclusion, which is required by the boundary conditions, to persist farther away from that inclusion as $\epsilon$ is decreased.  

 \section{Effective Membrane Viscosity}
 \label{sec:EffectiveViscosity}
  
 Armed with the results of the previous section, we now turn to computing the effective viscosity of a dilute suspension of membrane-bound inclusions.  As discussed above, we use the stress tensor to probe the dissipative processes in the system.  The effective membrane description of the suspension implies a coarse-graining of the system over length scales much larger than the size of the inclusions.  Thus, we compute the stress tensor averaged over the entire volume of the system $V_{\rm tot}$, which can be written as
\bb
\label{eq:AveStress}
\overline{\sigma}^{\rm tot}_{ij} \equiv \frac{1}{V_{\rm tot}} \int_{V_{\rm tot}} d^3 x \sigma^{\rm tot}_{ij} (\mathbf{x}).
\en
Due to the cylindrical symmetry of the problem, we choose $V_{\rm tot}$ to be a cylinder whose height $H_{\rm tot}$ and radius $R_{\rm tot}$ are large.  This volume includes the interiors of the solid inclusions. Within these regions, the stress tensor  $\sigma^{\rm tot}$ is not simply the fluid stress tensor; rather, it is the solid stresses in the inclusion caused by the fluids flows that surround it.  

Up until this point, we have been treating the membrane as a strictly two-dimensional, flat surface.  In this model, the stress within the membrane enters the integral above as a delta-function at the membrane surface $z=0$. It is convenient to avoid such a singularity when calculating the average stress tensor.  To do so, we use an equivalent \emph{three-dimensional} model of the membrane for which the stress is continuous at all points.  Specifically, a two-dimensional membrane with a viscosity $\eta_{\rm m}$ and two-dimensional pressure $p^{\rm m}$ is equivalent to a thin, three-dimensional fluid of thickness $h$, viscosity $\eta_{\rm m}/h$, and bulk pressure $p^{\rm m}/h$, in the limit of a vanishing membrane thickness $h \rightarrow 0$.~\cite{Saffman1, Saffman2, Hughes}  A schematic illustration of this 3D model is shown in Fig.~\ref{fig:Schematic2}.  Thus, we can compute the integrals in Eq.~(\ref{eq:AveStress}) using the three-dimensional model and then take the \emph{membrane limit} $h \rightarrow 0$ to recover the two-dimensional membrane considered in the previous Section. 

Consider first the membrane in the absence of the particulate suspension, with only the unperturbed flows $\mathbf{v}_0$ present.  From the results of Section~\ref{sec:flows}, it is straightforward to see that the stress tensor takes the form
\bb
\label{eq:StressZero}
\sigma_{0,ij} =  \delta_{\alpha i}^\perp \delta_{\beta j}^\perp \eta(z) c_{\alpha \beta}, \quad  \eta(z) \equiv 
 \begin{cases}
 \eta_- &z<0\\
 \frac{\eta_{\rm m}}{h} & 0<z<h\\
 \eta_+ &z>h
 \end{cases}.
 \en
Then the integral in Eq.~(\ref{eq:AveStress}) is given by 
 \bb
 \label{eq:AveStressZero}
 \overline{\sigma}_{0,ij} =  \delta_{\alpha i}^\perp \delta_{\beta j}^\perp \left[\left(\eta_+ + \eta_-\right) c_{\alpha \beta} +\frac{2}{H_{\rm tot}}\eta_{\rm m} c_{\alpha \beta}\right].
 \en

We now turn to the particulate suspension.  In the 3D membrane model, each inclusion is a solid cylinder whose height $h$ is equal to the membrane thickness; see Figure~\ref{fig:Schematic2}.  We anticipate that the average stress tensor for the suspension will have the same form as Eq.~(\ref{eq:AveStressZero}), with the membrane viscosity $\eta_{\rm m}$ being replaced by an effective membrane viscosity $\eta_{\rm m}^{\rm eff}$.  Thus, we \emph{define} the effective membrane viscosity via the average stress tensor:
\bb
\label{eq:etaEffDef}
 \overline{\sigma}^{\rm tot}_{ij} =  \delta_{\alpha i}^\perp \delta_{\beta j}^\perp \left[\left(\eta_+ + \eta_-\right) c_{\alpha \beta} +\frac{2}{H_{\rm tot}}\eta_{\rm m}^{\rm eff} c_{\alpha \beta}\right].
 \en

In order to calculate the average stress tensor for the suspension, we need the total velocity $\mathbf{v}^{\rm tot}$ for this system.  Since we work in the dilute limit, we can ignore the hydrodynamic interactions between the particles in the suspension.  That is, we discard the negligible alterations of the flow fields around one disk due to the presence of the other disks in the suspension, so that each disk is treated in isolation.  In this limit, the total fluid velocity is simply a linear superposition of the unperturbed membrane flows $\mathbf{v}_0$ and the perturbative flows from each disk in the suspension:
 \bb
 \label{eq:vTot}
 \mathbf{v}^{\rm tot} (r, \theta, z) = \mathbf{v}_0 (r, \theta, z) + \sum_{n=1}^N \mathbf{v}_1^{(n)} (r, \theta, z),
 \en
where $\mathbf{v}_1^{(n)} (r, \theta, z)$ is the perturbation to the flows $\mathbf{v}_0$ caused by an isolated disk whose center is located in the membrane at position $\mathbf{x}^{(n)}$, which can be obtained from Eq.~(\ref{eq:v1}) by a simple coordinate translation.  

Rather than attempting to directly compute the average stress Eq.~(\ref{eq:AveStress}) for the suspension, we first separate out the contributions of the unperturbed flows and of the perturbative flows of each particle in the suspension.  This can be accomplished by writing the average stress tensor as
 \bb
 \label{eq:AveStress1}
 \overline{\sigma}^{\rm tot}_{ij} = \overline{\eta(z) \partial_i v_j}+ \overline{\eta(z) \partial_j v_i}+\Xi_{ij},
 \en
 where
 \bb
\Xi_{ij} \equiv  \frac{1}{V_{\rm tot}} \int_{V_{\rm tot}} d^3 x \, \Big[ \sigma^{\rm tot}_{ij} -  \eta(z) \left(\partial_i v^{\rm tot}_j+ \partial_j v^{\rm tot}_i\right)\Big].
  \en

\begin{figure}
 \includegraphics[scale=0.8]{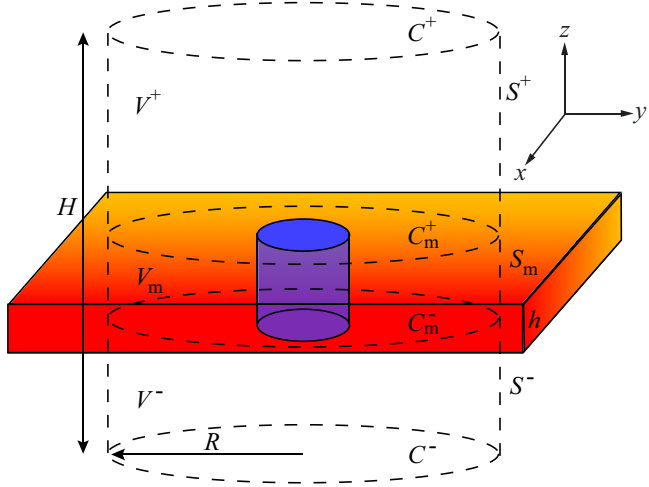}
 \caption{\label{fig:Schematic2} Schematic illustration of an isolated inclusion in a thin layer of fluid of thickness $h$, enclosed by a large cylinder of height $H$ and radius $R$.  The viscosity of the layer is $\eta_{\rm m}/h$; In the limit $h \rightarrow 0$, this system is equivalent to a two-dimensional membrane of viscosity $\eta_{\rm m}$.  The cylindrical volume $V$ is divided into volumes above ($V^+$), below ($V^-$), and within ($V_{\rm m}$) the fluid layer, as shown.  The cylindrical caps on these volumes are denoted by $C^\pm$ and $C_{\rm m}^\pm$, while the shells are denoted by $S^\pm$ and $S_{\rm m}$, as shown.}
 \end{figure}

Consider the first two terms in Eq.~(\ref{eq:AveStress1}).  Clearly, the contributions of the unperturbed flows to these terms will yield the unperturbed average stress tensor $\overline{\sigma}_{0, ij}$, Eq.~(\ref{eq:AveStressZero}).  Furthermore, we can show that the perturbative flows $\mathbf{v}_1^{(n)}$ do not contribute to these terms. Specifically, consider the quantity $\overline{\eta(z) \partial_i v^{(n)}_{1,j}}$. This clearly vanishes for $j=z$, but it also vanishes for $i=z$, because angular integration over an odd number of in-plane unit vectors $\hat{n}$ will vanish.  The $i=\alpha$ terms also vanish, because the integral evaluates to the velocity at $x_\alpha = \pm \infty$, where it vanishes.  Thus, the flows $\mathbf{v}_1^{(n)}$ do not contribute to the first two terms in Eq.~(\ref{eq:AveStress1}):
 \bb
 \label{eq:AveStress2}
 \overline{\sigma}^{\rm tot}_{ij} =  \overline{\sigma}_{0,ij}+\Xi_{ij},
 \en

We now turn to the integral $\Xi_{ij}$.  In the fluid regions of the system (i.e. outside of the rigid inclusions) the integrand is equal to the fluid pressure.  However, we know from the results of the previous section that this pressure vanishes everywhere outside of the membrane.  Furthermore, in the fluid regions of the membrane, we see from Eq.~(\ref{eq:PressureSolution}) that the angular dependence of the membrane pressure $\sim c_{\alpha \beta} n_\alpha n_\beta$.  Averaging over the angular variable $\theta$ produces the integral
 \bb
 \label{eq:nIntegral1}
 \frac{1}{\pi} \int_0^{2\pi} d\theta \, n_\alpha n_\beta = \delta_{\alpha \beta}.
 \en
From this we see that the contribution of the fluid membrane regions to $\Xi_{ij}$ also vanishes since $c_{\alpha \beta}$ is traceless.  Thus, the only regions of integration that contribute to $\Xi_{ij}$ are the solid interiors of the disks themselves.  Due to our neglect of the hydrodynamic interactions between the disks (as justified by the assumption of a dilute suspension), each disk in the membrane provides an identical contribution to $\Xi_{ij}$, so we have
 \bb
 \label{eq:Xi}
\Xi_{ij} = \frac{N}{H_{\rm tot} A_{\rm tot}} \int_V d^3 x \, \Big[\sigma_{1, ij}-  \eta(z) \left(\partial_i v_{1,j}+ \partial_j v_{1,i}\right)\Big],
 \en
where $ A_{\rm tot}=\pi R^2_{\rm tot}$ and $N$ is the number of particles in the suspension. The perturbative stress tensor $\sigma_{1, ij} \equiv \sigma_{ij}- \sigma_{0, ij}$, where, $\sigma_{ij}$ is the stress tensor everywhere within a system containing an isolated inclusion.  Thus, we have converted the computation of the average stress tensor of a particulate suspension into the problem of a single isolated inclusion considered in Section~\ref{sec:flows}.  Although the integrand is non-zero only within that inclusion, it proves useful to re-extend the region of integration $V$ to include all of the surrounding fluids.  Therefore, we choose $V$ to be a large cylinder whose height $H$ and radius $R$ will eventually be taken to infinity; see Fig~\ref{fig:Schematic2}.

Consider the first term in Eq.~(\ref{eq:Xi}), the integral of the perturbative stress tensor $\sigma_{1, ij}$.  From its definition, we can see that $\sigma_{1, ij}$ contains all of the solid stresses within the inclusion, as well as the fluid stresses caused by the perturbative velocity field $\mathbf{v}_1$.  Since the integration domain $V$ in Eq.~(\ref{eq:Xi}) clearly includes the interior of the solid inclusion, we would need to determine the solid stresses in this region to compute this integral directly.  We can avoid this difficulty, however, by using the divergence theorem to convert this volumetric integral into a surface integral.   Stress continuity requires that $\partial _k \sigma_{ik} =0$ at all points in space, including the interior of the inclusion. Furthermore, it is clear from Eq.~(\ref{eq:StressZero}) that  $\partial _k \sigma_{0,ik} =0$ everywhere.  Then we may write
\begin{align}
 \label{eq:StressInt1}
 \int_V d^3 x \, \sigma_{1, ij}  &=  \int_V d^3 x \, \partial_k \left(\sigma_{1, ik} x_j\right)\\
 \notag & = R \int_0^{2 \pi} d\theta \int_{-\infty}^\infty dz \, \Big[\sigma_{1,i\gamma} n_\gamma x_j\Big|_{r=R}.
 \end{align}
Here, we have extended the height $H$ of the enclosing cylinder to infinity. Because of the exponential decay of the perturbative fluid velocity Eq.~(\ref{eq:v1}) as $z \rightarrow \pm \infty$, we neglect the integration over its circular end-caps $C^\pm$ at $z=\pm H/2$ (see Fig.~\ref{fig:Schematic2}).  
 
 For $i=\alpha, j=z$ or $i=z, j=\alpha$, it is straightforward to show -- using Eq.~(\ref{eq:v1}) and the fact that $\partial_\alpha v_{1, \gamma}$ is even in $\hat{n}$ -- that the integrand of the surface integral in Eq.~(\ref{eq:StressInt1}) is odd in $\hat{n}$, and therefore vanishes upon integration over $\theta$.   For $i=j=z$, the integrand is $\propto c_{\gamma \beta} n_\gamma n_\beta$, which also vanishes upon integration over $\theta$, by Eq.~(\ref{eq:nIntegral1}).  Thus
\begin{align}
\label{eq:StressInt2}
 \int_V &d^3 x \, \sigma_{1, ij}  = \delta_{\alpha i}^\perp \delta_{\beta j}^\perp R^2 \int_0^{2 \pi} d\theta n_\gamma n_\beta \\
 \notag & \times \Bigg\{ \int_{-\infty}^\infty \eta(z)\left[\partial_\alpha v_{1,\gamma} + \partial_\gamma v_{1, \alpha} \right] dz- \delta_{\alpha \gamma} p_1^{\rm m}\Bigg|_{r=R} 
 \end{align}

The remaining terms in $\Xi_{ij}$ are proportional to the discontinuous viscosity function $\eta(z)$.  For these terms, we break up the integration volume $V$ into three different regions containing the three separate fluids in the system.  Namely, we divide $V$ into three separate cylinders $V^+, V^-,$ and $V^{\rm m}$, which enclose the regions $z>h$, $0<z<h$, and $z<0$, respectively; see Fig.~\ref{fig:Schematic2}. Using the divergence theorem, we obtain integrals of the velocity components $v_{1,\alpha}$ over the cylindrical shells $S^+, S^-$, and $S^{\rm m}$, whose outward normals are all $\hat{n}$, as well as integrals over the end-caps $C_{\rm m}^\pm$, whose outward normals are $\propto \hat{z}$.  Because the velocity is odd in $\hat{n}$ [see Eq.~(\ref{eq:v1})],  the latter integrals will all vanish.  Thus,
\begin{align}
\label{eq:VelocityInt}
\int_V &d^3 x \, \eta(z) \left(\partial_i v_{1,j}+ \partial_j v_{1,i}\right) \\
\notag &= \delta_{\alpha i}^\perp \delta_{\beta j}^\perp R \int_{-\infty}^\infty dz \, \eta(z) \int_0^{2\pi} \Big[n_\beta v_{1,\alpha} +n_\alpha v_{1, \beta}\Big]_{r=R}.
\end{align}

From Eqs.~(\ref{eq:v1}) and~(\ref{eq:p1}), we see that the $z$ integrals in Eqs.~(\ref{eq:StressInt2}) and~(\ref{eq:VelocityInt}) are all identical. Returning to the limit of an arbitrarily thin membrane, $h \rightarrow 0$, we find
\bb
\label{eq:zInt}
\lim_{h \rightarrow 0} \int_{-\infty}^\infty \eta(z) e^{- q \left| z \right|/a} = \eta_{\rm m} \left[ 1 + \frac{\epsilon}{q}\right].
\en

From Eqs.~(\ref{eq:StressInt2})--(\ref{eq:zInt}), we find that $\Xi_{ij} \propto \delta_{\alpha i}^\perp \delta_{\beta j}^\perp /H_{\rm tot}$.   Thus, the average stress tensor Eq.~(\ref{eq:AveStress2}) does indeed take the form of Eq.~(\ref{eq:etaEffDef}), as anticipated. Specifically, if we compute the remaining angular integrals in Eqs.~(\ref{eq:StressInt2}) and~(\ref{eq:VelocityInt}) using Eqs.~(\ref{eq:v1}) and~(\ref{eq:p1}), we find that the effective viscosity is
 \bb
 \label{eq:EffectiveViscosity1}
 \eta_{\rm m}^{\rm eff}= \eta_{\rm m}\left[1+ \phi \lim_{R \rightarrow \infty} \Phi \left(\frac{R}{a}\right) \right],
 \en
where $\phi \equiv N \pi a^2/A_{\rm tot}$ is the area fraction of particles in the membrane and 
\bb
\label{eq:Phi}
 \Phi (u) \equiv u^2 \int_0^\infty dq \, B(q) (q+\epsilon) \left[\frac{3}{8} J_2 (q u) - \frac{q u}{16} J_1 (qu)\right].
 \en 
In Eq.~(\ref{eq:EffectiveViscosity1}), we have taken the radius $R$ of the enclosing cylinder $V$ to infinity, as promised.    

We have succeeded in finding an expression for the effective membrane viscosity in terms of the function $B(q)$ determined in Section~\ref{sec:flows}.   Using Eq.~(\ref{eq:SingleIntSeries}), we may write Eq.~(\ref{eq:Phi}) as
\bb
\Phi (u) \equiv u^2 \sum_{n=1}^\infty b_n(\epsilon) \int_0^\infty dq \, j_{2n-1}(q)\left[\frac{3}{8} J_2 (q u) - \frac{q u}{16} J_1 (qu)\right].
 \en 
It is straightforward to show [see Eq.~(\ref{eq:BesselInt})] that the integral of the second term in brackets vanishes for all $n$, while the first integral vanishes for all $n>1$.  The $n=1$ integral is $\propto 1/u^2$, so its contribution to $\Phi(u)$ is $u$-independent and thus survives the $R \rightarrow \infty$ limit in Eq.~(\ref{eq:EffectiveViscosity1}). Thus, the effective membrane viscosity is determined solely from the coefficient $b_1(\epsilon)$:
 \bb
 \label{eq:EffectiveViscosity}
\eta_{\rm m}^{\rm eff}  =\eta_{\rm m}\left[1+ \phi \frac{b_1(\epsilon)}{4}\right] \equiv \eta_{\rm m} \left[ 1 + \phi f (\epsilon)\right].
 \en
 
The function $f(\epsilon)$ can be computed numerically for arbitrary values of $\epsilon$; see Appendix~\ref{app:Numerics}.  Before we discuss this solution, however, we first consider the asymptotic limits $\epsilon \gg 1$ and $\epsilon \ll 1$, where analytic solutions for $f(\epsilon)$ can be obtained.  

\subsection{$a \gg \ell_0$:  Large Inclusions}

When the particle size $a$ is much larger than the Saffman-Delbr\"{u}ck length $\ell_0$ -- that is, when $\epsilon \gg 1$ -- viscous dissipation occurs predominately in the surrounding 3D fluids, rather than in the membrane.  In this limit, we calculate the leading order and next-to-leading order dependence of $b_1 (\epsilon)$ on $\epsilon$.  We write
\bb
\label{eq:bLargeE}
b_n (\epsilon) \approx b_n^{(0)} + b_n^{(1)} (\epsilon).
\en
The leading order term $b_n^{(0)}$ is found by approximating $u + \epsilon \approx \epsilon$ in Eq.~(\ref{eq:MatrixElements}).  Using the orthogonality of the spherical Bessel functions Eq.~(\ref{eq:SphBesselID}), Eq.~(\ref{eq:bMatrixEq})  becomes
\bb
\sum_{n=1}^\infty b_n^{(0)} (\epsilon) \frac{\pi \delta_{l,n}}{2\epsilon (4l-1)}  = \frac{8}{3} \delta_{l,1}, \quad \Rightarrow \quad b_l^{(0)}(\epsilon) = \frac{16 \epsilon}{\pi} \delta_{l,1}.
\en

Using Eq.~(\ref{eq:bLargeE}), the next-to-leading order terms in Eq.~(\ref{eq:bMatrixEq}) are
\bb
\label{eq:bMatrixEqLargeEpsilon}
\sum_{n=1}^\infty \left[\tilde{b}_n^{(0)} \tilde{\mathcal{M}}_{n,l}^{(1)} (\xi)+ \tilde{b}_n^{(1)} (\xi) \tilde{\mathcal{M}}_{n,l}^{(0)}\right]  = 0,
\en
where $\tilde{b}_n = b_n/\epsilon$, $\xi \equiv 1/\epsilon$, $\tilde{\mathcal{M}}_{n,l}^{(0)}$ is given by Eq.~(\ref{eq:SphBesselID}), and 
\bb
\tilde{\mathcal{M}}_{n,l}^{(1)} (\xi) \equiv \lim_{\xi \rightarrow 0}\left[-\frac{1}{\xi} \int_0^\infty \frac{u du}{1+u} j_{2n-1}\left(\frac{u}{\xi}\right) j_{2l-1}\left(\frac{u}{\xi}\right)\right]. 
\en
The region $0<u<\xi$ of this integral gives a negligible contribution in the limit $\xi \rightarrow 0$ and can be discarded. In the remaining integral, we expand the spherical Bessel functions for large values of their arguments: $j_{2n-1} (x) \approx (-1)^n \cos (x)/x$ for $x \gg 1$. Using this approximation we find that the dominant contribution is logarithmic:  
\bb
\tilde{\mathcal{M}}_{n,l}^{(1)}(\xi) \approx \frac{1}{2} \left(-1\right)^{n+l} \xi \ln (\xi).
\en
Eq.~(\ref{eq:bMatrixEqLargeEpsilon}) must be satisfied for arbitrary values of $\xi$, as long as $\xi$ is sufficiently small.  As a result, the two terms in the sum of Eq.~(\ref{eq:bMatrixEqLargeEpsilon}) must have the same functional dependence on $\xi$. Then we find
\bb
\tilde{b}_n^{(1)} (\xi) = \frac{16 (4n-1)}{\pi^2} \left(-1\right)^n \xi \ln (\xi),
\en

Thus, in this limit
\bb
\label{eq:fLargeEps}
f (\epsilon) = \frac{4 \epsilon}{\pi}+ \frac{12}{\pi^2} \ln (\epsilon), \quad \epsilon \gg 1.
\en

As mentioned above, the flows in the bulk fluids dissipate much more energy than the flows in the membrane in the limit $\epsilon \gg 1$.  Therefore, it is more appropriate to define an effective \emph{three-dimensional} viscosity in this limit.  If we return to the average stress  Eq.~(\ref{eq:etaEffDef}) for a symmetric membrane ($\eta_+ = \eta_- \equiv \eta_{\rm 3D}$),
\bb
\overline{\sigma}_{ij} =2 \delta_{\alpha i}^\perp \delta_{\beta j}^\perp \eta_{\rm 3D} \left[1+\frac{2\ell_0}{H_{\rm tot}} + \frac{8 a \phi}{\pi H_{\rm tot}}\right]
\en
In the limit $\ell_0 \rightarrow 0$, the second term vanishes.  If we compare this to the average stress of the unperturbed membrane Eq.~(\ref{eq:AveStressZero}) in this limit, 
\bb
\overline{\sigma}_{ij} =2 \delta_{\alpha i}^\perp \delta_{\beta j}^\perp \eta_{\rm 3D},
\en 
 we can see that the effective 3-D viscosity is
 \bb
 \eta_{\rm 3D}^{\rm eff} = \eta_{\rm 3D} \left[1+\frac{8}{\pi} \phi_{\rm 3D}\right].
 \en
 where $\phi_{\rm 3D} \equiv N \pi a^3/V_{\rm tot}$ acts as a volume fraction of the particles.  This identification is not precise, since we have taken the $z$-extent of both the membrane and its inclusions to vanish in order to compute the effective membrane viscosity.  In writing $\phi_{\rm 3D}$ in this form, however, we have given the inclusions a vertical ($\hat{z}$) size of $a$ and neglected any numerical prefactors of order unity associated with the precise geometry of the inclusions (e.g. cylinders vs. spheres).  It is interesting to note that, in spite of this imprecision, the numerical prefactor $8/\pi \approx 2.55$ is within $2\%$ of the Einstein coefficient of $5/2$ for a three-dimensional suspension of spheres, Eq.~(\ref{eq:Eta3Deff}).  This result is not unexpected, despite the fact that there are profound differences in the underlying assumptions regarding the particle distribution between our calculation and that of Einstein.  In our model, all of the particles are confined to a plane, while in Einstein's work, the particles are assumed to occupy all space.  This distinction is lost, however, in taking the low volume fraction limit, where both calculations reduce to a \emph{single-particle} calculation.  In this essentially mean-field limit,  all hydrodynamic interactions between the particles are ignored, and the effective viscosity can depend on the mean volume fraction of the particles alone and not the details of their spatial distribution.  Thus, it is not surprising that our result closely approximates Einstein's when the dissipation in the membrane is negligible.
 
 \subsection{$a \ll \ell_0$:  Small Inclusions}
  
 We now consider the limit in which the Saffman-Delbr\"{u}ck length is large compared to the size of the inclusion, $a \ll \ell_0$.  We still assume that the suspension is dilute, so that $\ell_0$ is small compared to the mean lateral separation of the inclusions, $\ell_0 \ll a/\sqrt{\phi}$.  In this limit, we may continue to neglect the hydrodynamic interactions between the inclusions.  We now expect that the viscous dissipation occurs predominantly in the membrane.  
 
 In this limit, we expand
 \bb
 \frac{1}{u+\epsilon} \approx \frac{1}{u} -\frac{\epsilon}{u^2}, \qquad b_n(\epsilon) \approx b_n^{(0)} + \epsilon b_n^{(1)}.
 \en
 We require that Eq.~(\ref{eq:bMatrixEq}) be satisfied term by term in $\epsilon$, so that
 \bb
  b_n^{(0)} = \frac{8}{3} R_{1,n}^{-1},
  \en
  \bb
   b_n^{(1)} = \frac{8}{3} \sum_{l,n=1}^\infty R_{l,m}^{-1} R_{1,n}^{-1}\tilde{R}_{n,l},
   \en
   where the matrix elements $R_{n,l}$ and $\tilde{R}_{n,l}$ are given by Eqs.~(\ref{eq:R}) and~(\ref{eq:Rtilde}), respectively. The matrix inverse $R_{m,n}^{-1}$ can be computed analytically; see Ref.~\onlinecite{Hughes}.  The elements $R_{1,n}^{-1}$ are given by
   \bb
R_{1,n}^{-1}=\frac{3 \left(-1\right)^{n-1} (4n-1)\left(n-\frac{1}{2}\right)}{\pi n} \left(\frac{\Gamma\left(n-\frac{1}{2}\right)}{\Gamma\left(n\right)}\right)^2.
\en
Then $ b_n^{(0)} =  12$ and
 \bb
  b_n^{(1)} =\frac{8}{3} \sum_{l=1}^\infty R_{l,1}^{-1} \left[ \alpha \left(l+\tfrac{1}{2}\right)  R_{l+1,1}^{-1} + \alpha(l) R_{l,1}^{-1}\right] = \frac{32}{\pi}.
  \en
  
 Thus, in the limit $a \ll \ell_0$ (i.e. $\epsilon \ll 1$), we find
 \bb
 \label{eq:fSmallEps}
f (\epsilon) = 3 +\frac{8\epsilon}{\pi}, \qquad \epsilon \ll 1,
 \en
 so that, using Eq.~(\ref{eq:EffectiveViscosity}), 
 \bb
 \eta_{\rm m}^{\rm eff} = \eta_{\rm m} \left[1+ 3 \phi +\frac{8 a}{\pi \ell_0} \phi\right].
 \en
 The leading-order correction to the viscosity, which is reminiscent of the Einstein result Eq.~(\ref{eq:Eta3Deff}), gives the dependence of the effective membrane viscosity on the area fraction in the limit of vanishing inclusion radius.  The next-to-leading order contribution shows an additional viscosity enhancement, proportional to $a/\ell_0$.  This term is due to the additional dissipation caused by the flows induced in the bulk fluids for small (but finite) values of $a/\ell_0$.  
 
 It is important to recognize the distinction between the $\epsilon=0$ limit of our system and the strictly two-dimensional system of a suspension of infinite cylinders in a bulk fluid that has been considered previously.~\cite{Brady2, Belzons} Even when the viscosities of the surrounding bulk fluids vanish (i.e. $\ell_0 = \infty$), our system is not two-dimensional: The flows in the membrane extend infinitely far away from the membrane in the bulk fluids, but the pressure is still non-zero only within the membrane.  In contrast, the fluid pressure around an infinite cylinder is the same everywhere along the cylinder axis.  Thus, the $\epsilon=0$ limit of our system, where the viscosity correction $=3\phi$, is different than the viscosity correction of $2 \phi$ for a suspension of cylinders.~\cite{Brady2, Belzons}

\begin{figure}
 \includegraphics{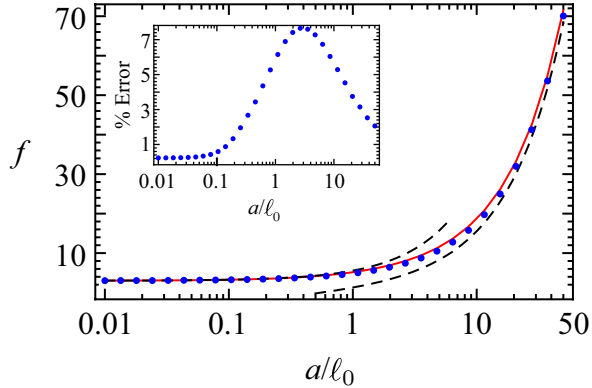}
 \caption{\label{fig:EtaEff} Exact numerical results (dots), asymptotic limits (dashed lines), and interpolation function (solid line) for the function $f(\epsilon)$.  For the numerical results, the infinite matrix is truncated at $6 \times 6$, and $Q = 5 \epsilon$; see Appendix~\ref{app:Numerics}.}
 \end{figure}

\subsection{Arbitrary $a/\ell_0$}
 
 While the above asymptotic results are suggestive, it is clearly desirable to examine $f(\epsilon)$ for arbitrary values of $\epsilon$.
 Figure~\ref{fig:EtaEff} shows the function $f(\epsilon)$ over several decades of $\epsilon$ values.  The exact numerical solution to Eq.~(\ref{eq:EffectiveViscosity}) is indicated by the points (see Appendix~\ref{app:Numerics} for details), while the asymptotic values Eqs.~(\ref{eq:fLargeEps}) and~(\ref{eq:fSmallEps}) are indicated by the dashed lines.  It is clear that the numerical solution agrees with the asymptotic expressions in the appropriate limits, and that the transition between these two limits is smooth and monotonic.  
 
 We can use the analytic expressions for the small and large $\epsilon$ behavior of $f (\epsilon)$ to construct an analytic function $\tilde{f}(\epsilon)$ that smoothly interpolates between these extremes.  We note from Eqs.~(\ref{eq:fLargeEps}) and~(\ref{eq:fSmallEps}) that $f (\epsilon)$ grows as $\epsilon + \ln(\epsilon)$ for large $\epsilon$ but has no logarithmic divergence at small $\epsilon$.  Thus, we need a term in the interpolation function $\tilde{f}(\epsilon)$  that grows logarithmically at large $\epsilon$ but remains well-behaved (i.e. non-singular) at small $\epsilon$.  The obvious choice is $\ln (1+\epsilon)$:
\bb
\tilde{f} (\epsilon) = \frac{12}{\pi^2} \ln \left(1+\epsilon\right) + g(\epsilon), 
\en
where
\bb
\label{eq:gLimits}
g(\epsilon)=
\begin{cases}
3 +\left(\frac{8}{\pi}-\frac{12}{\pi^2}\right) \epsilon &  \epsilon \ll 1\\
 \frac{4 \epsilon}{\pi} & \epsilon \gg 1\\
 \end{cases}.
 \en

To find a suitable function $g(\epsilon)$, we employ the method of two-point Pad\'{e} approximants:~\cite{Bender}
\bb
g(\epsilon) = \frac{A_N (\epsilon)}{C_M (\epsilon)},
\en
where $A_N$ and $C_M$ are polynomials of order $N,M$, respectively, in $\epsilon$.  Without loss of generality, we can set $C_M(0) =1$.  Therefore, we have $N+M+1$ unknown coefficients.  Three of these coefficients can be set by the known asymptotic limits of $g(\epsilon)$ given in Eq.~(\ref{eq:gLimits}).  Furthermore, in order to obtain $g(\epsilon) \sim \epsilon$ for $\epsilon \gg 1$, we must have $N=M+1$.  Thus, the first non-trivial Pad\'{e} approximant for $g(\epsilon)$ is $N=2, M=1$:
\bb
g(\epsilon) =\frac{a_0+a_1\epsilon + a_2 \epsilon^2}{1+c_1 \epsilon}.
\en
This has four unknown coefficients, so there is not a unique Pad\'{e} approximant for this function.  However, it is straightforward to show that $c_1$ is the undetermined coefficient, and that all values of $c_1>0$ give a smooth, monotonically increasing Pad\'{e} approximant, so we set $c_1=1$ for simplicity.  By expanding $g(\epsilon)$ for large and small values of $\epsilon$ and matching these limits to those given in Eq.~(\ref{eq:gLimits}), we find
\bb
\label{eq:Ftilde}
\tilde{f} (\epsilon) = \frac{12}{\pi^2} \ln \left(1+\epsilon\right) +\frac{3 \pi^2 + \left(3\pi^2+8\pi-12\right) \epsilon + 4 \pi \epsilon^2}{\pi^2\left(1+\epsilon\right)}.
\en

In Figure~\ref{fig:EtaEff}, we plot the interpolation function $\tilde{f}(\epsilon)$ as a solid line.  We see that it exhibits excellent agreement with the exact numerical results for all values of $\epsilon$.  Indeed, the error between $\tilde{f}(\epsilon)$ and the numerical solution never exceeds $8\%$, as shown in the inset of Fig.~\ref{fig:EtaEff}.

\section{Conclusion/Future Work}
\label{sec:Conclusion}

Membranes and fluid interfaces are by their nature hybrid systems.  Although the membrane/interface is itself two-dimensional, it is surrounded by bulk three-dimensional fluids.  As a result, the hydrodynamics of membranes can exhibit both two-dimensional and three-dimensional characteristics, depending on the system in question.  The effective viscosity of a fluid membrane containing rigid inclusions demonstrates this \emph{dimensional crossover}.  For particles whose radii $a$ are small compared to the Saffman-Delbr\"{u}ck length $\ell_0$, the effect of the suspension on the large length scale viscous dissipation under shear can best be thought of as providing an increase in the effective viscosity of the membrane $\eta_{\rm m}^{\rm eff}$.  Conversely, for large inclusions relative to $\ell_0$, their effect on the sheared membrane and surrounding solvent can be understood as an increase in the viscosities of the bulk solvents that is proportional to the \emph{volume fraction} of the inclusion. For arbitrary inclusion size, we have determined a reasonably simple interpolation formula that gives an accurate estimate of the exact numerical solution for the effective membrane viscosity.  

Mathematically, we have seen that the hydrodynamics of membranes containing rigid inclusions is a mixed boundary value problem whose solution obeys a set of dual integral equations.   One of the benefits of this work is that it helps to elucidate the mathematically machinery needed to solve these dual integral equations.  Since these equations arise in many problems in membrane hydrodynamics, we plan to use our newfound understanding of the mathematics to solve other problems.  Despite the mathematical complexity of the problem presented here, it is in a way one of the simpler problems one can consider in the hydrodynamics of membranes with rigid inclusions, because the hydrodynamic interactions between the inclusions can be ignored.  Indeed, one of the motivations for this work was to study a system in which we could learn about membrane hydrodynamics with rigid inclusions without the additional complication of particle interactions.  Armed with this knowledge, we plan to investigate such interparticle interactions in future studies.    One problem of particular biophysical significance is the study of lubrication forces between two large membrane inclusions in close proximity.  It is now widely believed that many transmembrane proteins recruit lipid rafts~\cite{Simons} in the cell's plasma membrane.  Treating these extended structures as essentially rigid objects, one may ask how the hydrodynamic interactions between two such rafts affect the kinetics of protein aggregation in the membrane.  Similar questions can also be asked of the kinetics of phase separation in the multicomponent lipid bilayers of giant unilamellar vesicles.~\cite{Veatch}  More generally we expect the mathematical and physical features of the problem considered in this paper to arise in the study of the kinetics of inclusions or finite size domains in any lipid bilayer or Langmuir monolayer system.  Such problems should exhibit the phenomenon of a scale-dependent dimensional crossover explored here.  

MLH and AJL thank H. Stone for interesting conversations.  This work was supported in part by grant NSF-CMMI0800533.

\appendix

\section{Dual Integral Equations}
\label{app:DualIntegral}

In this Appendix, we present the mathematical tools necessary to manipulate the integral equations, Eqs.~(\ref{eq:IntEqIn})--(\ref{eq:IntEqOut}), and then use these tools to derive a single integral equation.   The necessary integral operator identities are presented here without proof; we refer the reader to Ref.~\onlinecite{Sneddon} for the derivation of these identities. 

Consider a function $f(q)$ defined everywhere on the positive-$q$ axis, $0<q<\infty$.  Adopting the compact notation used in Ref.~\onlinecite{Sneddon}, we denote the modified Hankel transform of this function by the operator $S_{\eta, \lambda} f(u)$, which is defined by
\bb
\label{eq:HankelDef}
S_{\eta, \lambda} f(u)\equiv \frac{2^\lambda}{u^\lambda} \int_0^\infty  dq \frac{J_{2\eta+\lambda} (q u)}{q^{\lambda-1}}  f(q) .
\en
Where necessary we use the expanded notation $S_{\eta, \lambda} f(u) = S_{\eta, \lambda} \left\{f(q);u\right\}$.  Using this notation, Eqs.~(\ref{eq:IntEqIn}) and~(\ref{eq:IntEqOut}) can be written as, respectively,
\bb
\label{eq:IntEqInSneddon}
S_{0,2} B(u) = 4, \quad u<1,
\en
\bb
\label{eq:IntEqOutSneddon}
S_{\frac{3}{2},-1} B(u) = -\frac{2}{\epsilon x} S_{2,-2} B(u), \quad u>1.
\en
We will return to the final integral equation, Eq.~(\ref{eq:IntEqIn2}), at the end of this Appendix.  

The principal difficulty presented by Eqs.~(\ref{eq:IntEqInSneddon}) and~(\ref{eq:IntEqOutSneddon}) is that the unknown function $B(u)$ is defined by two separate integral equations, each with its own domain of applicability.    From the closure relation for Bessel functions, we know that that the inverse of a modified Hankel transform is another modified Hankel transform: specifically, $S^{-1}_{\eta, \lambda} = S_{\eta+\lambda,-\lambda}$.  Therefore, the inversion of a modified Hankel transform is possible only if it appears in an equation that applies to the entire half-line $0<u<\infty$.   The dual integral equations, Eqs.~(\ref{eq:IntEqInSneddon}) and~(\ref{eq:IntEqOutSneddon}), clearly do not satisfy this requirement.   As a result, we cannot directly invert the modified Hankel transforms in Eqs.~(\ref{eq:IntEqInSneddon}) and~(\ref{eq:IntEqOutSneddon}) to solve for the function $B(u)$.

To resolve this dilemma, we combine Eqs.~(\ref{eq:IntEqInSneddon}) and~(\ref{eq:IntEqOutSneddon}) into a single integral equation using the Erd\'{e}lyi-Kober operators defined below.  Through the application of these operators, we can write the left-hand side of Eqs.~(\ref{eq:IntEqInSneddon}) and~(\ref{eq:IntEqOutSneddon}) \emph{in the same form}.  In this way, we generate a single integral equation whose domain of validity extends over the entire real positive axis.  

The Erd\'{e}lyi-Kober operators are defined as
\bb
\label{eq:IDef}
I_{\eta, \lambda} f(q) = \frac{2 q^{-2\lambda-2\eta}}{\Gamma(\lambda)} \int_0^q \frac{\left(q^2-u^2\right)^{\lambda-1}}{u^{-2\eta-1}} f(u) \, du,
\en
\bb
\label{eq:KDef}
K_{\eta, \lambda} f(q) = \frac{2 q^{2\eta}}{\Gamma(\lambda)} \int_q^\infty \frac{\left(u^2-q^2\right)^{\lambda-1}}{u^{2\lambda+2\eta-1}} f(u) \, du.
\en
These integrals only converge for $\lambda>-1/2$; for $\lambda<-1/2$, 
\bb
\label{eq:ILambdaNeg}
I_{\eta, \lambda} f(q) =q^{-2\eta-2\lambda-1} \mathcal{D}_q^n\left[\frac{I_{\eta,\lambda+n} f(q)}{q^{-2\eta-2\lambda-2n-1}}\right],
\en
\bb
\label{eq:KLambdaNeg}
K_{\eta, \lambda} f(q) =(-1)^n q^{2\eta-1} \mathcal{D}_q^n\left[\frac{K_{\eta-n,\lambda+n} f(q)}{q^{2\eta-2n-1} }\right],
\en
where $n$ is an integer such that $\lambda+n>0$ and 
\bb
\mathcal{D}_q f(q) \equiv \frac{1}{2} \frac{\partial}{\partial q} \left(\frac{f(q)}{q}\right).
\en

The utility of these operators stems from the following observations:  (i) When acting on a function $f(q)$, the operators  $I_{\eta, \lambda}$ and $K_{\eta,\lambda}$ involve integrals over $(0,q)$ and $(q,\infty)$, respectively. Thus, they depend on two disjoint subspaces of the positive real line.  This is essential because it allows one to apply $I_{\eta, \lambda}$ to Eq.~(\ref{eq:IntEqInSneddon}) and obtain an integral that is well-defined for $0<u<1$.  Similarly, we can apply $K_{\eta, \lambda}$ to Eq.~(\ref{eq:IntEqOutSneddon}) and obtain an integral that is well-defined for $1<u<\infty$.  (ii)  Both operators  $I_{\eta, \lambda}$ and $K_{\eta, \lambda}$ have the simple convolution properties with modified Hankel transforms, namely
\bb
\label{eq:ItimesS}
I_{\eta+\lambda,\gamma} S_{\eta,\lambda}  =S_{\eta,\lambda+\gamma}
\en
and
\bb
\label{eq:KtimesS}
K_{\eta,\lambda} S_{\eta+\lambda,\gamma}  = S_{\eta,\lambda+\gamma}.
\en
We now apply $I$ to Eq.~(\ref{eq:IntEqInSneddon}) and $K$ to Eq.~(\ref{eq:IntEqOutSneddon}), making a judicious choice of the coefficients so that the left-hand sides of the resulting equations are identical.  Specifically, it is straightforward to show using Eqs.~(\ref{eq:ItimesS}) and~(\ref{eq:KtimesS}) that 
 \bb
 I_{2,-\frac{3}{2}} S_{0,2} = K_{0,\frac{3}{2}} S_{\frac{3}{2},-1} = S_{0,\frac{1}{2}}.
 \en
 The coeffients of these operators are unique and set by the coefficients in Eqs.~(\ref{eq:IntEqInSneddon}) and~(\ref{eq:IntEqOutSneddon}), along with the requirement that the resultant modified Hankel transform be the same in both equations.  Thus, Eqs.~(\ref{eq:IntEqInSneddon}) and~(\ref{eq:IntEqOutSneddon}) can be written as a single integral equation.
 \bb
 S_{0,\frac{1}{2}} B(u) = 
 \begin{cases}
 I_{2,-\frac{3}{2}} (4) &u<1\\
 -\frac{2}{\epsilon} K_{0,\frac{3}{2}}\left[\frac{1}{u} S_{2,-2} B(u) \right]& u>1\\
 \end{cases}.
 \en
 From Eqs.~(\ref{eq:IDef}) and~(\ref{eq:ILambdaNeg}), we can show that
 \bb
 I_{2,-\frac{3}{2}} (4) = \frac{\mathcal{D}_u^2\left[u^6 I_{2,\frac{1}{2}} (4)\right]}{u^2}  =\frac{ \mathcal{D}_u^2\left[\frac{64u^6}{15 \sqrt{\pi}}\right]}{u^2}= \frac{16}{\sqrt{\pi}}.
 \en
Using the identity
 \bb
\label{eq:KProp}
K_{\eta, \lambda}\left[u^{2\gamma} f(u)\right] =u^{2\gamma} K_{\eta-\gamma,\lambda} f(u)
\en
along with Eq.~(\ref{eq:KtimesS}), we find
\bb
 K_{0,\frac{3}{2}}\left[\frac{1}{u} S_{2,-2} B(u) \right]=\frac{1}{u} K_{\frac{1}{2},\frac{3}{2}} S_{2,-2} B(u) = \frac{1}{u} S_{\frac{1}{2},-\frac{1}{2}} B(u).
  \en
  Finally, by applying the inverse Hankel transform $S^{-1}_{0,\frac{1}{2}} = S_{\frac{1}{2},-\frac{1}{2}}$, the single integral equation reduces to
    \bb
  \label{eq:SingleIntSneddon}
B(u) = S_{\frac{1}{2},-\frac{1}{2}} \left.
 \begin{cases}
\frac{16}{\sqrt{\pi}} &w<1\\
 -\frac{2}{\epsilon w} S_{\frac{1}{2},-\frac{1}{2}} B(w)& w>1\\
 \end{cases} ; u\right\} 
 \en

We can write Eq.~(\ref{eq:SingleIntSneddon}) as a conventional integral equation using Eq.~(\ref{eq:HankelDef}). Noting that $J_{1/2} (w u) = \sqrt{2/(\pi w u)} \sin (w u)$, we find
 \begin{align}
  \label{eq:SingleInt1}
B(u) = &\frac{16}{\pi}\left[\frac{\sin u}{u^2}-\frac{\cos u}{u}\right]\\
\notag&-\frac{2}{\epsilon\pi} \int_0^\infty dz \, z B(z) \int_1^\infty dw \sin(w u) \sin (w z).
\end{align}
Writing $\sin(w u) \sin (w z)= \cos[w(u-z)]-\cos[w(u+z)]$ and noting that $\int_0^\infty dw \cos(w y) = \pi \delta(y)$, we perform the integration over $w$, 
  \begin{align}
  \label{eq:SingleInt2}
   \pi \left(u+\epsilon\right) &B(u) = 16 \epsilon \left[\frac{\sin u}{u^2}-\frac{\cos u}{u}\right]\\
  \notag &
+  \int_0^\infty dz \, z B(z)\left[\frac{\sin(u-z)}{u-z} -\frac{\sin(u+z)}{u+z} \right].
 \end{align}
 It is this form of the single integral equation that we use in Section~\ref{sec:flows} to solve for the function $B(u)$.   

Finally, we return to the third boundary condition, Eq.~(\ref{eq:IntEqIn2}), which to this point we have neglected.  For this integral equation the kernel is $J_3 (q u)$.  It is straightforward to show that 
\bb
\int_0^\infty dq  J_3 (q u)\left(\frac{\sin q}{q^2}-\frac{\cos q}{q}\right)=0.
\en
Similarly, for all $w>1$,
\bb
\int_0^\infty dq J_3 (q u)  \sin (w q)=0.
\en
From these two integrals we note that Eq.~(\ref{eq:IntEqIn2}) is automatically satisfied by any solution $B(u)$ of Eq.~(\ref{eq:SingleInt1}); that is, this condition provides no unique information about the function $B(u)$.

\section{Numerics}
\label{app:Numerics}

In order to obtain a numerical solution for the coefficients $b_n (\epsilon)$, we need to compute the integrals $\mathcal{M}_{n,l}$ defined in Eq.~(\ref{eq:MatrixElements}).  However, the oscillatory nature of the spherical Bessel functions makes these integrals difficult to compute numerically.  We can avoid this difficult by dividing the integration region into two portions:  $0<q<Q$ and $Q <q < \infty$, where $Q \gg \epsilon$.  Then in the latter region we perform a Taylor expansion of the denominator.  Keeping terms up to second order in $1/q$,
\begin{align}
\notag
\mathcal{M}_{n,l} \approx& \int_0^Q dq \, j_{2n-1} (q) j_{2l-1} (q)\left[\frac{1}{q+\epsilon}-\frac{1}{q}+\frac{\epsilon}{q^2}\right]\\
&+R_{n,l} - \epsilon \tilde{R}_{n,l},
\end{align}
where~\cite{Hughes}
\begin{align}
\label{eq:R}
R_{n,l} &\equiv \int_0^\infty \frac{dq}{q} j_{2n-1}(q) j_{2l-1} (q) \\
\notag
&= \frac{\left(-1\right)^{n+l-1}}{2\left(4 \left(n-l\right)^2-1\right)(n+l-1)(n+l)},
\end{align}
\begin{align}
\label{eq:Rtilde}
\tilde{R}_{n,l} &\equiv \int_0^\infty \frac{dq}{q^2} j_{2n-1}(q) j_{2l-1} (q) \\
\notag
&=\frac{1}{2}\alpha \left(l+\tfrac{1}{2}\right) \delta_{n,l+1} +\alpha (l) \delta_{n,l} + \tfrac{1}{2}\alpha \left(l-\tfrac{1}{2}\right) \delta_{n,l-1}, 
\end{align}
with
\bb
\alpha (l) \equiv \frac{\pi}{(4l+1)(4l-1)(4l-3)}.
\en
The remaining integral in $\mathcal{M}_{n,l}$, whose integration region is $0<q<Q$, can be computed numerically, as long as $\epsilon$ is not too large ($\epsilon \lesssim 50$).  This procedure yields accurate values for these integrals for sufficiently large values of $Q$; throughout this paper, we use $Q=10 \epsilon$. 

To compute the membrane velocity Eq.~(\ref{eq:v1Solution}), we proceed in the same manner:
\begin{align}
\notag
\mathcal{V} (u;n,m)\approx& \int_0^Q dq \, j_{2n-1} (q) J_m (q u)\left[\frac{1}{q+\epsilon}-\frac{1}{q}+\frac{\epsilon}{q^2}\right]\\
&+\mathcal{I}(u;n,m,1) -\epsilon \mathcal{I}(u;n,m,2),
\end{align}
where for $u>1$ and $-\frac{3}{2}<p<m+2n$,
\begin{align}
\label{eq:BesselInt}
\mathcal{I} &(u;n,m,p) \equiv \int_0^\infty \frac{dq}{q^p} \, j_{2 n-1} (q) J_m (qu)\\
\notag
&=\frac{\sqrt{\pi} \, \Gamma\left(n+\frac{m-p}{2}\right) F\left(n+\frac{m-p}{2}, n-\frac{m+p}{2}; 2n+\frac{1}{2}; \frac{1}{u^2}\right)}{2^{p+1} u^{2n-p} \Gamma\left(2n+\frac{1}{2}\right) \Gamma\left(\frac{m+p}{2}-n+1\right)},
\end{align}
and $F(\alpha, \beta;\gamma;z)$ is the hypergeometric function. It is very important to note that this integral vanishes if $\frac{1}{2}(m+p)-n+1$ is a non-positive integer, due to the divergence of the Gamma function in the denominator.

Finally, in order to determine the coefficients $b_n (\epsilon)$, the infinite matrix $\mathcal{M}_{n,l}$ must be inverted.  To do this inversion numerically, we truncate the matrix.  It is straightforward to verify that accurate solutions for the velocities and effective viscosity are obtained for reasonably-sized matrices; for all of the numerical results presented in this paper, we truncate the matrix at $10 \times 10$.

 \end{document}